\def\nk{n_{\rm b}}

\def\rfr#1{Equation~\ref{#1}}

\def\virg#1{``#1"}

\def\eqi{\begin{equation}}
\def\eqf{\end{equation}}
\def\eqia{\begin{eqnarray}}
\def\eqfa{\end{eqnarray}}

\def\rp#1#2{{#1\over#2}}
\def\lb#1{\label{#1}}

\def\ton#1{\left(#1\right)}

\documentclass{aastex}

\usepackage{float}
\usepackage{amsmath,textgreek,w-greek,wasysym}
\usepackage{amscd,lineno}
\usepackage{amssymb,pbox}
\usepackage{graphicx,epsfig}
\usepackage{txfonts}
\RequirePackage{color}

\newcommand{\emaila}{lorenzo.iorio@libero.it}

\allowdisplaybreaks[1]

\begin{document}

\title{Are we close to \textcolor{black}{putting}  the anomalous perihelion precessions from Verlinde's emergent gravity to the test?}

\shortauthors{L. Iorio}

\author{Lorenzo Iorio\altaffilmark{1} }
\affil{Ministero dell'Istruzione, dell'Universit\`{a} e della Ricerca
(M.I.U.R.)-Istruzione
\\ Permanent address for correspondence: Viale Unit\`{a} di Italia 68, 70125, Bari (BA),
Italy}
\email{\emaila}

\begin{abstract}
In the framework of the emergent gravity scenario by Verlinde, it was recently observed by Liu and Prokopec that, among other things, an anomalous pericenter precession would affect the orbital motion of a test particle orbiting an isolated central body. Here, it is shown that, if it were real, its expected magnitude for the inner planets of the Solar System would be at the same level of the present-day accuracy in constraining any possible deviations from their standard perihelion precessions as inferred from long data records spanning about the last century. The most favorable situation for testing the Verlinde-type precession seems to occur for Mars. Indeed, according to recent versions of the EPM and INPOP planetary ephemerides, non-standard perihelion precessions, of whatsoever physical origin, which are larger than some $\approx 0.02-0.11$ milliarcseconds per century are not admissible, while the putative precession predicted by Liu and Prokopec amounts to $0.09$  milliarcseconds per century. Other potentially interesting astronomical and astrophysical scenarios like, e.g., the Earth's LAGEOS II artificial satellite, the double pulsar system PSR J0737-3039A/B and the S-stars orbiting the Supermassive Black Hole in Sgr A$^\ast$ are, instead, not viable because of the excessive smallness of the predicted effects for them.
\end{abstract}

%\centerline
%{Keywords: Oort Cloud--Kuiper belt: general--celestial mechanics--ephemerides}

%

keywords{
gravitation---celestial mechanics--ephemerides
}
\section{Introduction}
In the framework of the novel emergent gravity theory proposed by \citet{2016arXiv161102269V} to account for the Dark Matter phenomenology,
\citet{2016arXiv161200861L} observed that, among other things, a test particle orbiting an isolated  body of mass $M$ would experience an additional pericenter shift per orbit
\eqi\Delta\varpi_\textrm{theo}=\pi\sqrt{\rp{2\mu H_0}{3c^3}},\lb{shift}\eqf
where $\varpi$ is the longitude of pericenter\footnote{\textcolor{black}{It is a broken, \virg{dogleg} angle, defined as \citep{2000ssd..book.....M} $\varpi=\omega+\Omega$, where $\omega,~\Omega$ are the argument of pericenter and the longitude of the ascending node, respectively.}},  $\mu\doteq GM$ is the gravitational parameter of the central object, $G$ is the Newtonian gravitational constant, $c$ is the speed of light in vacuum, and $H_0$ is the Hubble parameter. From \rfr{shift}, it is useful to infer an analytical expression for a secular pericenter precession which can  be conveniently compared with recent observational determinations  of such an orbital effect like, e.g., the supplementary perihelion precessions of some planets of the Solar System \citep{2011CeMDA.111..363F,2013MNRAS.432.3431P,2016arXiv160100947F}. Since the orbital period of a test particle is
\eqi P_\textrm{b}=\rp{2\pi}{\nk}=2\pi\sqrt{\rp{a^3}{\mu}},\eqf where $\nk$ is the Keplerian mean motion and $a$ is the semimajor axis of the orbit, it turns out
\eqi\dot\varpi_\textrm{theo} = \rp{\Delta\varpi_\textrm{theo}}{P_\textrm{b}} =\mu\sqrt{\rp{H_0}{6 a^3 c^3}}.\lb{peritheo}\eqf
It is important to note that if, on the one hand, \rfr{peritheo} is proportional to $\mu$, on the other hand, it falls off as \textcolor{black}{\eqi \dot\varpi_\textrm{theo}\propto a^{-3/2}\lb{dep}\eqf} for a given value of the central mass; thus, the more the particle is close to its parent star, the larger its expected extra-precession is. \textcolor{black}{Moreover, \rfr{dep} is a distinctive feature of the modified model of gravity considered here which would be of great help, in principle, in discriminating Verlinde's theory from the anomalous precessions due to other more or less viable alternative models \citep{2007PhRvD..75h2001A,2007PhRvD..75f4011A,2009isft.confE..18I} by taking, e.g, the ratio of the perihelion rates for a pair of planets \citep{2009isft.confE..18I,2015IJMPD..2430015I}.}

The paper is organized as follows. Section \ref{periplan} deals with the inner planets of the Solar System by comparing their theoretically predicted precessions $\dot\varpi_\textrm{theo}$ as per \rfr{peritheo} with the most recent determinations $\Delta\dot\varpi_\textrm{obs}$ of the supplementary perihelion precessions from planetary ephemerides. In Section \ref{altri}, other potentially appealing astronomical and astrophysical scenarios are investigated: the terrestrial artificial satellite LAGEOS II, the double pulsar PSR J0737-3039A/B, and the S-stars orbiting the Supermassive Black Hole in the center of the Galaxy. Section \ref{fine} is devoted to our conclusions.
\section{The perihelion precessions of the planets of the Solar System}\lb{periplan}
The last decades have witnessed remarkable advances in the field of  planetary orbit determination for almost all of the major bodies of the Solar System in view of the availability of increasingly long records of radiotechnical data (ranges and range-rates) collected by a host of spacecraft targeted to many of them. As a consequence, any possible deviations from standard, i.e. general relativistic and Newtonian, perihelion precessions are nowadays tightly constrained for several planets.

Table \ref{periV} shows the latest determinations $\Delta\dot\varpi_\textrm{obs}$ of the admissible ranges for whatsoever possible anomalous  perihelion precessions  obtained by independent teams of astronomers with different planetary ephemerides.
\begin{table*}
\caption{The column $\dot\varpi_\textrm{theo}$ collects the theoretically predicted perihelion precessions, calculated for some planets of the Solar System according to \rfr{peritheo} with $H_0 = 67.6~\textrm{km}~\textrm{s}^{-1}~\textrm{Mpc}^{-1}$ \citep{2016arXiv160703143G}. The columns $\Delta\dot\varpi_\textrm{obs}$ display the supplementary perihelion precessions observationally determined for the same planets with the planetary ephemerides EPM2011 \citep{2013MNRAS.432.3431P}, INPOP10a \citep{2011CeMDA.111..363F}, INPOP15a \citep{2016arXiv160100947F} by reducing almost one century of data of many different types with fully relativistic dynamical and measurement models; the Verlinde-type effect was not modeled in any of the ephemerides considered. They should be regarded as representative of the ranges of admissible values for any possible anomalous planetary perihelion precessions induced by whatsoever non-standard dynamical effect: they are all statistically compatible with zero. \textcolor{black}{Recently, \citet{ParkAJ2017} constrained the secular precession of the orbital element $\omega +\Omega\cos I$ of Mercury to an accuracy of $1.5$~mas~cty$^{-1}$. About Saturn, \citet{2014PhRvD..89j2002H}, in looking for possible anomalous effects on it due to a certain version of MOND, managed to constrain its perihelion precession to the $0.43$~mas~cty$^{-1}$ level.} All the figures quoted here for the precessions, both theoretical and experimental,  are in milliarcseconds per century (mas cty$^{-1}$). For the benefit of the reader, we converted the results by \citet{2013MNRAS.432.3431P} from milliarcseconds per year (mas yr$^{-1}$) to mas cty$^{-1}$. }
\label{periV}
\centering
\begin{tabular}{lllll}
\noalign{\smallskip}
\hline
& $\dot\varpi_\textrm{theo}$ & $\Delta\dot\varpi_\textrm{obs}$ (EPM2011) & $\Delta\dot\varpi_\textrm{obs}$ (INPOP10a) &  $\Delta\dot\varpi_\textrm{obs}$ (INPOP15a)\\
\hline
Mercury & $0.7$ & $-2.0\pm 3.0$ & $0.4\pm 0.6$ & $0.0\pm 2.1$\\
Venus & $0.3$ & $2.6\pm 1.6$ & $0.2\pm 1.5$ & $-$ \\
Earth & $0.2$ & $0.19\pm 0.19$ & $-0.2\pm 0.9$ & $-$  \\
Mars & $0.09$ & $-0.020\pm 0.037$ & $-0.04\pm 0.15$ & $-$ \\
\textcolor{black}{Jupiter} & \textcolor{black}{$0.014$} & \textcolor{black}{$58.7\pm 28.3$} & \textcolor{black}{$-41\pm 42$} & \textcolor{black}{$-$} \\
Saturn & $0.006$ & $-0.32\pm 0.47$ & $0.15\pm 0.65$ & $0.6\pm 2.6$  \\
\hline
\end{tabular}
\end{table*}
From a comparison with the analytically computed Verlinde-type perihelion rates $\dot\varpi_\textrm{theo}$ of \rfr{peritheo}, listed in Table \ref{periV} as well, it can be noted that their magnitudes are very close to the present-day level of accuracy in constraining the planetary apsidal precessions from observations\textcolor{black}{, at least as far as the inner planets are concerned. Instead, the orbits of the gaseous giants Jupiter and Saturn, which, in principle, might be regarded as potentially interesting targets because of the relative smallness of other competing precessions induced by otherwise accurately modeled \textit{N}-body perturbations, are known with insufficient accuracies with respect to \rfr{peritheo}. Given that the uncertainties quoted in Table \ref{periV} for Saturn already come from analyses of multi-year radiotechnical data from the Cassini spacecraft, it is unlikely that even the reduction of the entire data record of this probe, which, after more than 12 years in Saturn orbit, will end its mission in September 2017 by plunging through the planet's atmosphere \citep{2016DPS....4851401S}, will ameliorate  the situation so much to allow for a detection of the Verlinde-type precession of the ringed giant}.

\textcolor{black}{According to Table \ref{periV}, the most favorable situation seems to occur, at present, for Mars. Indeed, its predicted theoretical perihelion precession amounts to \eqi \dot\varpi_\textrm{theo} = 0.09~\textrm{milliarcseconds~per~century}~\ton{\textrm{mas~cty}^{-1}},\lb{theoM}\eqf while the accuracy in constraining any possible anomalous precession from observations is  \eqi \sigma_{\Delta\dot\varpi_\textrm{obs}}= 0.037~\textrm{mas~cty}^{-1}\eqf if the EPM2011 ephemerides are considered \citep{2013MNRAS.432.3431P}. Actually, }
the EPM2011-based supplementary rate \citep{2013MNRAS.432.3431P} would even rule out the expected emergent anomalous effect at a $3-\sigma$ level since \textcolor{black}{\eqi\rp{\left|\dot\varpi_\textrm{theo}-\Delta\dot\varpi^\textrm{best}_\textrm{obs}\right|}{\sigma_{\Delta\dot\varpi_\textrm{obs}}}=3.\eqf} For all the other planets, and also for Mars if the supplementary rate by the INPOP10a ephemerides is considered,  the Verlinde-type precessions $\dot\varpi_\textrm{theo}$ are compatible with the admissible ranges $\Delta\dot\varpi_\textrm{obs}$. \textcolor{black}{Recently, \citet{ParkAJ2017} considered the in-plane precession\footnote{\textcolor{black}{\citet{ParkAJ2017} use the symbol $\varpi$ for the in-plane, unbroken angle $\omega+\Omega\cos I$.}} $\dot\omega + \dot\Omega\cos I$ of the pericenter of Mercury by constraining it to the $1.5$~mas~cty$^{-1}$ level. It is still $2.1$ times larger than the Verlinde-type in-plane precession which coincides with \rfr{peritheo} since the node is not affected by the modified model of gravity considered here. }

\textcolor{black}{In view of the potential opportunity offered by Mars, it is important to assess as accurately as possible the main systematic errors which may plague such a measurement.}

\textcolor{black}{It has been known for a long time \citep{1984Icar...57....1W,2009A&A...508..479M, 2010P&SS...58..858S,2013Icar..222..243K} that a major source of perturbation for the orbit of Mars is represented by the asteroids crowding in the belt located between the orbits of the red planet and Jupiter. It is so because of the uncertainty still affecting our knowledge of some of their key physical parameters, mainly their masses. Luckily, recent advances in modeling their dynamical action allowed for a sufficiently accurate determination of their overall mass for our purposes. Indeed, \citet{2016IAUS..318..212P}, by means of the EPM2014 ephemerides, modeled the ensemble of such perturbers, including both the individual 301 largest asteroids and the minor ones, as a planar annulus with internal and external radii equal to $2.06$ au and $3.27$ au, respectively, being able to determine its total mass as \eqi m_\textrm{belt} = (12.25\pm 0.19)\times 10^{-10}~M_\odot.\eqf In their Table 3, \citet{2016IAUS..318..212P} worked out the nominal perihelion advance of Mars for ten revolutions due to an annulus with the aforementioned radii and $m=10^{-10}~M_\odot$: it amounts to \eqi \Delta\varpi_\textrm{10~rev} = 0.497285998~\textrm{mas}.\eqf It is straightforward to infer a mismodeled perihelion precession as little as \eqi\sigma_{\dot\varpi_\textrm{belt}} = 0.04~\textrm{mas~cty}^{-1},\lb{aster}\eqf which is $2.25$ times smaller than the expected Verlinde-type precession. Incidentally, \rfr{aster} explicitly shows  how the asteroids, in fact, are just the dominant source of uncertainty in constraining the perihelion precession of Mars.
}

\textcolor{black}{Another potential source of major systematic uncertainty is represented by the \textit{N}-body perturbations due to the other planets, especially Jupiter and Saturn. We will, now, demonstrate that, luckily,  they can be modeled with sufficiently high accuracy. The perihelion precession induced by a distant pointlike perturber P is proportional to \citep{1962GeoJ....6..271C}
\eqi\dot\varpi_\textrm{P}\propto \rp{\mu_\textrm{P}}{a^3_\textrm{P}\nk},\lb{periP}\eqf where $\nk$ is the Keplerian mean motion of the disturbed planet.
The present-day uncertainty in the Jupiter's mass is \citep{2016P&SS..126...78L} \eqi\sigma_{\mu_\textrm{Jup}}=1.5\times 10^9~\textrm{m}^3~\textrm{s}^{-2},\eqf which corresponds to a relative accuracy of \eqi \rp{\sigma_{\mu_\textrm{Jup}}}{\mu_\textrm{Jup}} = 1\times 10^{-8}.\eqf Thus, \rfr{periP} tells us that the
mismodelled perihelion precession of Mars due to Jupiter is of the order of \eqi\sigma_{\dot\varpi_\textrm{Jup}}\lesssim 0.02~\textrm{mas~cty}^{-1};\eqf it is $4.5$ times smaller than \rfr{theoM}. The ongoing Juno mission will allow, among other things, for a remarkable improvement in our knowledge of $\mu_\textrm{Jup}$ \citep{2016P&SS..126...78L}.
According to \citet{2006AJ....132.2520J}, the current uncertainty in the mass of Saturn is \eqi\sigma_{\mu_\textrm{Sat}}=1.1\times 10^9~\textrm{m}^3~\textrm{s}^{-2},\eqf corresponding to a relative accuracy of \eqi \rp{\sigma_{\mu_\textrm{Sat}}}{\mu_\textrm{Sat}} = 3\times 10^{-8}.\eqf Thus, according to \rfr{periP}, the resulting mismodelled perihelion precession of Mars due to Saturn is of the order of \eqi\sigma_{\dot\varpi_\textrm{Sat}}\lesssim 0.002~\textrm{mas~cty}^{-1},\eqf which is negligible with respect to \rfr{theoM}.
}

\textcolor{black}{The stationary component of the general relativistic field of the Sun, yielding the gravitomagnetic Lense-Thirring effect \citep{1918PhyZ...19..156L} due to the solar angular momentum $S_\odot$, has not yet been modeled  in all the modern ephemerides; thus, in principle, it may act as a further source of systematic uncertainty. In a coordinate system with its fundamental plane aligned with the Sun's equator, the nominal Lense-Thirring perihelion precession of Mars amounts to \citep{1918PhyZ...19..156L} \eqi\dot\varpi_\textrm{LT} = \rp{2G S\ton{1-3\cos I}}{c^2 a^3\ton{1-e^2}^{3/2}} = -0.03~\textrm{mas~cty}^{-1},\lb{dodtLT}\eqf whose magnitude is $3$ times smaller than \rfr{theoM}. In \rfr{dodtLT}, $e,~I$ are the orbital eccentricity and the inclination\footnote{\textcolor{black}{The inclination of the orbital plane of Mars to the Sun's equator is $I = 5.65~\textrm{deg}$.}} of the orbital plane to the equator of the Sun, respectively. }

\textcolor{black}{A further potential systematic bias is represented by the competing perihelion precession \citep{2005som..book.....C}
 \eqi \dot\varpi_{J_2} = \rp{3\nk R^2 J_2\ton{3-4\cos I +5\cos 2 I}}{8a^2\ton{1-e^2}^2},\lb{dodtJ2}\eqf
 due to the Sun's quadrupole mass moment $J_2$ which was recently  determined by \citet{ParkAJ2017} to $4\%$  as \eqi J_2=\ton{2.25\pm 0.09}\times 10^{-7}.\eqf In \rfr{dodtJ2}, which holds  in a coordinate system whose fundamental plane is aligned with the Sun's equator, $R$ is the solar equatorial radius. It turns out that the mismodeled perihelion precession of Mars due to the Sun's oblateness is
\eqi\sigma_{\dot\varpi_{J_2}}= 0.008~\textrm{mas~cty}^{-1},\eqf which is about 10 times smaller than \rfr{theoM}. }

\textcolor{black}{The static, Schwarzschild-like component of the general relativistic field of the Sun, which is fully modeled in the softwares of the astronomers who produce the planetary ephemerides, yields a nominal gravitoelectric perihelion precession of Mars as large as \eqi\dot\varpi_\textrm{GE} = \ton{\rp{2 + 2\gamma - \beta}{3}}\rp{3\nk\mu}{c^2 a\ton{1-e^2}} =1350.93~\textrm{mas~cty}^{-1}.\lb{dodtGE}\eqf In \rfr{dodtGE}, $\gamma,~\beta$ are the most important PPN parameters, which are equal to 1 in general relativity.
 The actual mismodeling in \rfr{dodtGE} is due to the uncertainties in $\gamma,~\beta$. The most accurate determination of $\gamma$ is currently at the  \citep{2003Natur.425..374B} \eqi\sigma_\gamma = 2\times 10^{-5}\eqf level, while $\beta$ was recently determined with an accuracy \citep{ParkAJ2017} \eqi\sigma_\beta = 4\times 10^{-5}.\eqf Thus, the resulting uncertainty in the gravitoelectric perihelion precession of Mars is \eqi \sigma_{\dot\varpi_\textrm{GE}}=0.025~\textrm{mas~cty}^{-1},\eqf which is $3.6$ times smaller than \rfr{theoM}.
}

\textcolor{black}{Table \ref{errors} summarizes the competing precessions of the perihelion of Mars. }
\begin{table*}
\caption{ Main unmodeled/mismodeled  perihelion rates of Mars, in mas cty$^{-1}$, acting as systematic errors for a potential measurement of the Verlinde-type precession $\dot\varpi_\textrm{theo}=0.09~\textrm{mas}~\textrm{cty}^{-1}$. }
\label{errors}
\centering
\begin{tabular}{lll}
\noalign{\smallskip}
\hline
Dynamical effect & Uncertainty in the key parameter(s) & Perihelion precession  \\
\hline
Asteroid belt & $\sigma_{m_\textrm{belt}}=1.9\times 10^{-11}~M_\odot$  \protect{\citep{2016IAUS..318..212P}} & $0.04$\\
Lense-Thirring & Nominal  &  $-0.03$\\
Schwarzschild & \pbox{20cm}{$\sigma_\gamma=2\times 10^{-5}$ \protect{\citep{2003Natur.425..374B}}  \\ $\sigma_\beta=4\times 10^{-5}$ \protect{\citep{ParkAJ2017}} } & $0.025$ \\
Jupiter attraction & $\sigma_{\mu_\textrm{Jup}} = 1.5\times 10^9~\textrm{m}^3~\textrm{s}^{-2}$ \protect{\citep{2016P&SS..126...78L}} & $0.02$\\
Solar oblateness & $\sigma_{J_2}=9\times 10^{-9}$  \protect{\citep{ParkAJ2017}} & $0.0086$ \\
Saturn attraction & $\sigma_{\mu_\textrm{Sat}} = 1.1\times 10^9~\textrm{m}^3~\textrm{s}^{-2}$ \protect{\citep{2006AJ....132.2520J}}& $0.002$\\
\hline
\end{tabular}
\end{table*}

However, caution is in order since dedicated, covariance-based, \textcolor{black}{global} data reduction \textcolor{black}{for all the major bodies of the Solar System for which accurate radio-tracking is available should be performed by explicitly modeling the exotic gravitational dynamics considered here. That is in order} to avoid, or, at least, reduce, the risk that part of the putative anomalous signature, if really existent, may be removed from the post-fit residuals being partly absorbed in the estimated values of, say, the planetary initial conditions. On the other hand, our analysis shows that the perspectives of effectively putting the Verlinde-type emergent gravity to the test in the Solar System seems promising, especially if Mars is considered in view of the steady improvements in its orbit determination allowed by the accurate tracking of the remarkable number of present and future landers and orbiters \textcolor{black}{\citep{2011Icar..211..401K}. Furthermore, also the orbit of Mercury should be improved in the next years thanks to the steady analysis of the entire data record of the MESSENGER spacecraft, of which the work by \citet{ParkAJ2017} is a first step\footnote{\textcolor{black}{For an earlier analysis based on shorter data record, see \citet{2014A&A...561A.115V}.}}, and the future BepiColombo mission \citep{2010P&SS...58....2B, 2016Univ....2...21S}, scheduled to be launched in 2018.}

\textcolor{black}{Finally, as we will see in the next Section,} it does not seem plausible that other astronomical and astrophysical systems can become viable test-beds for the emergent gravity scenario considered here in a foreseeable future.
\section{Other astronomical and astrophysical scenarios}\lb{altri}
Indeed, as far as the Earth's laser ranged LAGEOS II satellite is concerned, \rfr{peritheo} returns a perigee precession as little as $0.007$ milliarcseconds per year (mas yr$^{-1}$) for it. Suffice it to say that the standard, Schwarzschild-like, general relativistic perigee precession of LAGEOS II, which is larger than the Verlinde-type precession by a factor of $10^6$, was recently measured with a claimed accuracy of about $2\%$ \citep{2010PhRvL.105w1103L,2014PhRvD..89h2002L}.

Also the double pulsar system, PSR J0737-3039A/B \citep{2003Natur.426..531B,2004Sci...303.1153L}, may not be frutifully used. Indeed, while the current accuracy in determining its periastron precession amounts to about $2.5$ arcseconds per year (arcsec yr$^{-1}$)  \citep{2006Sci...314...97K}, \rfr{peritheo} tells us that its expected Verlinde-type precession is just $0.01$ arcsec yr$^{-1}$.

Unfortunately, also the stellar system orbiting the Supermassive Black Hole in the Galactic center \citep{2009ApJ...692.1075G,2008ApJ...689.1044G} does not offer better perspectives. Indeed, according to \rfr{peritheo}, the expected anomalous periastron precession of the S-star S0-102 \citep{2012Sci...338...84M} is $31$ mas cty$^{-1}$; it is hopelessly smaller than the standard general relativistic precession $\dot\varpi_\textrm{GR} = 32.5~\textrm{arcsec}~\textrm{yr}^{-1}$, which has not yet been measured, by a factor of $10^{-6}$.
\section{Conclusions}\lb{fine}
Thus, it seems that, as far as the anomalous pericenter precession is concerned, the emergent gravity scenario by Verlinde may not be put to the test anywhere in any foreseeable future with the notable exception of the Solar System's planetary motions. Indeed, the current accuracy in constraining any possible anomalous perihelion rates from the latest planetary ephemerides is at the same level, or perhaps even better for Mars, of the expected Verlinde-type precessions for the inner planets. Just as an example, the predicted anomalous perihelion precession of Mars amounts to $0.09$ milliarcseconds per century, while the EPM2011 ephemerides are able to constrain any possible deviations from standard (prograde) perihelion precessions  to the $0.017$ milliarcseconds per century level. Further improvements in the orbit determination of Mercury are expected from the approved BepiColombo mission, while also the knowledge of the  orbit of Mars will be steadily improved in view of the continuous tracking of an increasing number of landers and spacecraft. \textcolor{black}{A global fit of purposely modified dynamical models including the Verlinde-type gravity to extended records of accurate radiotechnical measurements for several major bodies of the Solar System would be an important step towards the practical implementation of such an ambitious goal.}

\bibliography{PXbib,IorioFupeng}{}
%-----------------------------------------

\end{document}